\documentclass[12pt]{article}
\usepackage{amsmath}
\usepackage{amsfonts}
\usepackage{amssymb}
\usepackage{amstext}
\usepackage{graphicx}
\usepackage{epsfig}%
\usepackage{graphicx}%
\usepackage{color}

\title{\centerline
\bf Lee-Wick radiation induced bouncing universe models}
\author{ Kaushik Bhattacharya$^{1}$, Yi-Fu Cai$^{2}$, and Suratna Das$^{3}$,
\thanks{email:\,\,\, $^{1}$kaushikb@iitk.ac.in, \,\,\,$^{2}$yifucai@physics.mcgill.ca,\,\,\, $^{3}$suratna@tifr.res.in}
\\
\normalsize
$^{1}$Department of Physics, Indian Institute of Technology, Kanpur,
\\
\normalsize
Kanpur 208016, India\\
\normalsize
$^{2}$Department of Physics, McGill University, Montr\'eal, QC H3A 2T8, Canada
\\
\normalsize
$^{3}$Tata Institute of Fundamental Research, Homi Bhabha Road,\\
\normalsize
Colaba, Mumbai 400005, India
}
\begin{document}
\maketitle
\begin{abstract}
The present article discusses about the effect of a Lee-Wick partner
infested radiation phase of the early universe. As Lee-Wick partners
can contribute negative energy density so it is always possible that
at some early phase of the universe when the Lee-Wick partners were
thermalized the total energy density of the universe became very small
making the effective Hubble radius very big. This possibility gives
rise to the probability of a bouncing universe. As will be shown in
the article that a simple Lee-Wick radiation is not enough to produce
a bounce. There can be two possibilities which can produce a bounce in
the Lee-Wick radiation phase. One requires a cold dark matter
candidate to trigger the bounce and the other possibility requires the
bouncing temperature to be fine tuned such as all the Lee-Wick
partners of the standard fields are not thermalized at the bounce
temperature. Both the possibilities give rise to blue-tilted power
spectrum of metric perturbations. More over the bouncing universe
model can predict the lower limit of the masses of the Lee-Wick
partners of chiral fermions and massless gauge bosons. The mass limit
intrinsically depends upon the bounce temperature.
\end{abstract}

\section{Introduction}

Recently the idea of constructing a unitary, gauge-invariant,
Lorentz-invariant and divergence free theory of Quantum
Electrodynamics by invoking unusual partners of the standard particles
in the Lagrangian, originally proposed by Lee and Wick
\cite{Lee:1969fy, Lee:1970iw}, has been generalized to construct a
Lee-Wick theory of the standard model of particle physics
\cite{Grinstein:2007mp}. This model, while still suffers from
unsettled issues such as quantum instability of the vacuum due to
gravitational interaction \cite{Cline:2003gs}, attracts a lot of
attention in the literature for several phenomenological
benefits. Namely, the mass of the Higgs field in such a quantum field
theory is perturbatively stable against the quadratically divergent
radiative corrections and thus the well-known ``hierarchy puzzle" of
the Standard Model of particle physics could be avoided. There are
many other interesting applications of the Lee-Wick idea of which some
are discussed here. In \cite{Carone:2008bs} a minimal extension of
Lee-Wick Standard model (LWSM) is considered, in \cite{Carone1} a LWSM
with more than one Lee-Wick (LW) partner for each standard model
particle is studied.  Ref.~\cite{Carone2} deals with gauge-coupling
unification in Lee-Wick framework and in \cite{Carone3,
  Alvarez:2011ah} analysis of two-Higgs doublet models where one of
the doublet contains Lee-Wick fields is presented. In
Ref. \cite{Krauss:2007bz} the process $gg\rightarrow h_0\rightarrow
\gamma\gamma$ is studied in the framework of LWSM where the authors
predict small changes in the rate of these processes, due to the
presences of Lee-Wick fields, from those rates calculated from other
models such as universal extra dimensions. In \cite{Figy:2011yu} Higgs
pair production processes $gg\rightarrow h_0h_0$ and $gg\rightarrow
h_0 \tilde p_0$ are studied in LWSM framework.

It was recently realized in \cite{Cai:2008qw} that a model constructed
out of a Lee-Wick type scalar field theory is capable of yielding a
bounce during the evolution of the universe within the framework of a
homogeneous and isotropic cosmological background.  In this model the
energy density and pressure of all other kinds of fields are supposed
to be negligible. In this scenario the Lee-Wick partners, which arise
from higher derivative operators, evolve as a tracking solution to the
normal matter (the standard scalar field) and can break certain energy
conditions when the energy scale of the universe becomes high
enough. Therefore, a non-singular bounce takes place which avoid the
big bang singularity widely existing in the standard FRW
cosmologies. Interests in such Lee-Wick non-singular bouncing models
have increased recently as these models \cite{Wands:1998yp,
  Finelli:2001sr} can be considered as an alternative to inflation and
can be used to explain the origin of large scale structure in the
universe. In this paradigm the primordial vacuum fluctuations leaves
the Hubble radius during a matter-dominated contracting phase and then
form a nearly scale-invariant power spectrum after the bounce.

However, it was found that non-singular Lee-Wick bouncing cosmologies
induced by pressure-less matter are in general unstable while
accompanied with radiation and a bouncing solution might be achieved
only when an extremely fine tuning of the initial phases of the field
configuration is assumed \cite{Karouby:2010wt, Karouby:2011wj}. This
problem can be roughly understood as follows. The normal matter which
scales as $a^{-3}$ as a function of the scale factor $a$ could be
easily suppressed by radiation matter which scales as $a^{-4}$ in the
contracting phase. If we simply add a radiation component to
non-relativistic matter within a framework of a Lee-Wick bounce model
without fine tuning of the initial values of radiation then the
radiation would become dominant, during the contraction phase of the
universe, and produce a big crunch singularity before the bounce takes
place. In Ref. \cite{Karouby:2010wt}, the effects of Lee-Wick partners
of radiation fields were taken into account as well but the energy
scale of the universe is assumed to be below the scale of
thermalization of Lee-Wick radiation. Then, it was found the effective
energy density of Lee-Wick radiation increases slower than that of
regular radiation and thus cannot help to realize the bounce at low
temperature regime.

But unlike these models of bouncing cosmologies, where the energy
density and pressure of the universe arises from a scalar field (whose
energy scales as $a^{-3}$) or a radiation field (whose energy scales
as $a^{-4}$), one can conceive a phase of the early universe where we
have the normal standard model particles and their Lee-Wick parters
(as virtual resonances) in thermal equilibrium as shown in
Refs. \cite{Fornal:2009xc, Bhattacharya:2011bb}. This phase is a
non-standard radiation dominated phase where one can have a state
parameter $w \sim 1$ and consequently the energy density scales $\sim
a^{-6}$. It is explicitly described in Ref. \cite{Bhattacharya:2011bb}
that in such kind of a Lee-Wick partner infested radiation dominated
universe the time-temperature relation in standard cosmology changes
and the universe may go through out-of equilibrium processes when some
of the Lee-Wick partners become non-relativistic.  It was noticed that
these kind of cosmologies can produce negative energy densities and
pressure when one includes chiral fermions and massless gauge bosons
\cite{Wise:2009mi, Bhattacharya:2012te}. This particular point turned
out to be a serious drawback for these models. It was shown that only
under some specific conditions these kind of an early radiation
domination can exist. On the other hand bouncing cosmologies naturally
require a time (bounce time) when the Hubble parameter turns out to be
zero (indicating a zero energy density). It turns out that one can
construct an interesting model of a bouncing cosmology from the
non-standard Lee-Wick radiation phase.

In this paper we study the possibility of realizing a Lee-Wick bounce
at high temperature by taking into account the radiation fields. The
Lee-Wick resonances are produced dramatically when the temperature of
the radiation fluid is much higher than the mass of heavy Lee-Wick
fields. As the contribution of Lee-Wick radiation to the total energy
density is always negative it enlarges the viable phase space for the
universe to experience the nonsingular bounce. Unlike the method
developed in \cite{Karouby:2010wt} in which the radiation fields are
described by the field configuration of gauge fields, we directly
study thermodynamics of normal particles and their Lee-Wick partners
following \cite{Fornal:2009xc, Bhattacharya:2011bb}. However, for a
radiation dominated universe purely dominated by standard fields and
their Lee-Wick partners, one still cannot obtain a nonsingular
bounce. This is because, at the bouncing moment when the energy
density of Lee-Wick resonances are able to cancel that of normal
radiation, the pressure of the universe keeps positive and then
prevent the background to evolve into an expanding phase. Thus we
introduce a cold dark matter (CDM) component in our model which can
easily solve this problem. As a consequence, our cosmological model,
which realizes a nonsingular bouncing solution in the thermal Lee-Wick
theory, consists of cold dark matter fluid, radiation, and Lee-Wick
resonances.

If one does not include the CDM component then one cannot obtain the
required condition of the bounce as discussed in the last
paragraph. This fact is true if all the standard particles and their
Lee-Wick partners are in thermal equilibrium near the bounce time. But
if it happens that the bounce temperature is such that some of the
Lee-Wick partner's mass are more than the bounce temperature then the
plasma around the bounce time will have less Lee-Wick
partner. Interestingly, this kind of a model where the bounce
temperature is less than some of the masses of the Lee-Wick partners
can give rise to a non-singular bounce. This model does not require
any CDM candidate to make the pressure negative at the bounce time. In
the present article we will also discuss about this alternative
bouncing model. Both of the bouncing models yields a blue-tilted power
spectrum of metric perturbations in the expanding phase after bounce.

The outline of this paper is as follows: In Section 2 we briefly
introduce the thermodynamics of Lee-Wick particles. In Section 3, we
study the condition for a Lee-Wick thermal bounce and then provide a
concrete cosmological model which involves both kind of bounce
scenarios as described above. Section 4 is devoted to the discussion
of cosmological perturbations in both of these models. The final
section presents some conclusions and discussion. For completeness we
have attached an appendix at the end of the article which discusses
about the mode matching conditions near the bounce point.

\section{Thermodynamics of Lee-Wick particles revisited}

In this section we will give a brief account of the thermodynamics of
Lee-Wick particles. We will show that in a realistic model of Lee-Wick
theory, where all the degrees of freedom of the standard particles
along with their Lee-Wick partners are taken into account properly,
the energy density of the Lee-Wick resonance dominated universe would become
very small when the heaviest of the Lee-Wick partners  thermalize
due to the very high temperature of the universe. This property is in
favor of a bouncing scenario which requires the Hubble parameter
(which is directly proportional to the square-root of the total energy
density of the universe) to vanish at the time of bounce.

It is noted in \cite{Fornal:2009xc, Bhattacharya:2011bb} that the
Lee-Wick particles behave very differently from their standard
partners in the relativistic regime. For the standard bosons and
fermions the energy density, pressure and the entropy density read as
\begin{eqnarray}
\rho^{(\rm sm)}_b= \frac{g\pi^2 T^4}{30}\,,\,\,\,
p^{(\rm sm)}_b= \frac{g\pi^2 T^4}{90}\,,\,\,\,
s^{(\rm sm)}_b= \frac{2g\pi^2 T^3}{45}\,,
\label{smb}
\end{eqnarray}
and
\begin{eqnarray}
\rho^{(\rm sm)}_f= \frac{7g\pi^2 T^4}{240}\,,\,\,\,
p^{(\rm sm)}_f= \frac{7g\pi^2 T^4}{720}\,,\,\,\,
s^{(\rm sm)}_f= \frac{7g\pi^2 T^3}{180}\,,
\label{smf}
\end{eqnarray}
respectively. Here the subscripts $b$ and $f$ stands for bosons and
fermions and $g$ stands for any number of internal degree of freedom
of the relativistic species. On the other hand, it is shown in
\cite{Fornal:2009xc, Bhattacharya:2011bb} that the Lee-Wick particles
at high energies contribute negatively to the energy density, pressure
and entropy density as
\begin{eqnarray}
\rho^{(\rm LW)}_b&=& -g\left(\frac{\pi^2 T^4}{30} - \frac{M^2
    T^2}{24}\right)\,,
\label{lwenb}\\
p^{(\rm LW)}_b&=&-g\left(\frac{\pi^2 T^4}{90} - \frac{M^2
    T^2}{24}\right)\,,
\label{lwpb}\\
s^{(\rm LW)}_b&=&-g\left(\frac{2\pi^2 T^3}{45} - \frac{M^2 T}{12}\right)
\label{lwsb}
\end{eqnarray}
for the bosonic Lee-Wick partners and
\begin{eqnarray}
\rho^{(\rm LW)}_f&=& -g\left(\frac{7\pi^2 T^4}{240} - \frac{M^2
    T^2}{48}\right)\,,
\label{lwenf}\\
p^{(\rm LW)}_f&=&-g\left(\frac{7 \pi^2 T^4}{720} - \frac{M^2
    T^2}{48}\right)\,,
\label{lwpf}\\
s^{(\rm LW)}_f&=&-g\left(\frac{7\pi^2 T^3}{180} - \frac{M^2 T}{24}\right)
\label{lwsf}
\end{eqnarray}
for the fermionic Lee-Wick partners. In the above equations $M$ is the
mass of a generic Lee-Wick partner and as the system is relativistic
$T \gg M$. As the Lee-Wick particles were initially introduced as
resonance particles to overcome the divergences appearing in a theory
and consequently when $T \gg M$ the standard model particles are
naturally ultrarelativistic.  If one considers a toy model where each
standard particle is accompanied by one Lee-Wick partner and the
number of degrees of freedom of the Lee-Wick partner is same as that
of its standard partner then in such a scenario the total energy
density, pressure and entropy density turn out to be positive as
\cite{Fornal:2009xc, Bhattacharya:2011bb}
\begin{eqnarray}
\rho_b&=& \rho^{(\rm sm)}_b + \rho^{(\rm LW)}_b = \frac{g M^2 T^2}{24}\,,
\label{nenb}\\
p_b&=& p^{(\rm sm)}_b + p^{(\rm LW)}_b = \frac{g M^2 T^2}{24}\,,
\label{npb}\\
s_b&=& s^{(\rm sm)}_b + s^{(\rm LW)}_b= \frac{g M^2 T}{12}
\label{nsb}
\end{eqnarray}
for the bosonic sector and
\begin{eqnarray}
\rho_f&=& \rho^{(\rm sm)}_f + \rho^{(\rm LW)}_f = \frac{g M^2 T^2}{48}\,,
\label{nenf}\\
p_f&=& p_f^{(\rm sm)} + p_f^{(\rm LW)} = \frac{g M^2 T^2}{48}\,,
\label{npf}\\
s_f&=& s_f^{(\rm sm)} + s_f^{(\rm LW)}= \frac{g M^2 T}{24}
\label{nsf}
\end{eqnarray}
for the fermionic sector. But this simple scenario, where each
standard particle is accompanied by one Lee-Wick particle with equal
number of degrees of freedom, does not serve the purpose while dealing
with realistic scenarios. It is discussed in \cite{Wise:2009mi} that
in the fermionic sector each chiral fermions require two Lee-Wick
partners to eliminate the higher derivative terms in a initial higher
derivative Lagrangian to construct a Lee-Wick theory. Such
discrepancies also arise in bosonic sector when one considers massive
Lee-Wick partners (with 3 degrees of freedom) of massless gauge bosons
(with two longitudinal degrees of freedom)
\cite{Bhattacharya:2012te}. Both these cases lead to unacceptable
scenarios with negative energy density. It is shown in
\cite{Bhattacharya:2012te} that considering both the fermionic and
bosonic sector together one still can come up with a scenario where
the total energy density, pressure and entropy density can become
positive.

In a realistic scenario, where each fermion is accompanied by two
Lee-Wick partners and the extra degrees of freedom of the massive
Lee-Wick partners of each massless gauge boson are taken into account,
the total energy density of the realistic Lee-Wick plasma can be
written as \cite{Bhattacharya:2012te}
\begin{eqnarray}
\rho=\frac{\tilde{M}^2}{24}\tilde{g}_{*N}T^2 - \frac{7\pi^2}{240}
\tilde{g}_FT^4 - \frac{\pi^2}{30}nT^4.
\label{energyd2}
\end{eqnarray}
Here the new number of degrees of freedom $\tilde{g}_{*N}$ is given as
\begin{eqnarray}
\tilde{g}_{*N}= \sum_{i={\rm bosons}}
g_{iN}\left(\frac{M_i}{\tilde{M}}\right)^2
\left(\frac{T_i}{T}\right)^2
+ \sum_{i={\rm fermions}} g_{iF} \left(\frac{M_i}{\tilde{M}}\right)^2
\left(\frac{T_i}{T}\right)^2\,,
\label{lwgstar2}
\end{eqnarray}
where $g_{iN}$ for bosonic particles stands for the number of internal
degrees of freedom $g_i$ for the partners of massive standard bosons
(may be 2 or 1), while for standard massless vector boson partners it
equals $g_i+1$ where primarily $g_i=2$. Also the unpaired fermionic
contribution comes with $\tilde{g}_F$ where
\begin{eqnarray}
\tilde{g}_F = \sum_{i={\rm fermions}} g_{iF} \left(\frac{T_i}{T}\right)^4\,,
\label{lwgf}
\end{eqnarray}
where $\tilde{g}_F$ solely arises from the unpaired fermionic Lee-Wick
partners of the standard model particles. The quantity $n$ is defined
as
\begin{eqnarray}
n= \sum_{i={\rm massive\,vect.\,bosons}}
\left(\frac{T_i}{T}\right)^4\,.
\label{newn}
\end{eqnarray}
Here $T_i$ is the temperature at which the $i^{\rm th}$ massive
Lee-Wick vector boson partner of a standard massless gauge boson is
equilibrated and $n$ denotes the number of massive vector boson
partners of massless standard gauge bosons if all the species are in
thermal equilibrium at the same temperature $T$. The sum appearing in
Eq. (\ref{newn}) does not include all the massive Lee-Wick vector
boson partners, but includes only those which are partners of massless
standard gauge bosons. The total pressure of these Lee-Wick infested
plasma is given by
\begin{eqnarray}
p=\frac{\tilde{M}^2}{24}\tilde{g}_{*N}T^2 -
\frac{7\pi^2}{720}\tilde{g}_FT^4 - \frac{\pi^2}{90}nT^4\,.
\label{pres2}
\end{eqnarray}
It must be noted that if the energy density of the Lee-Wick partner
infested universe is positive then the pressure of the same universe
must be positive. It is also to be noted that the negative
contributions to the total energy density and pressure are coming from
the Lee-Wick partners of the chiral fermions and massless gauge
bosons.

\section{Model(s) of Lee-Wick thermal bounce}

Our main aim is to construct a non-singular bouncing cosmology within
the arena of thermal Lee-Wick scenario in which the universe initially
starts its evolution in a contracting phase and evolves into a
standard thermal expanding phase smoothly and continuously through a
non-singular bounce. We first start with a general discussion on the
conditions required for a non-singular bounce and would consider a
simple model where the universe consists of bosonic and fermionic
radiation fields and their thermalized Lee-Wick resonances. We will
show that this simple toy model, which consists of only radiation
fields accompanied with their thermal Lee-Wick partners, is incapable
of achieving the conditions required for a non-singular bounce and
then we will try to illustrate two possible modifications to this
simple model where a bounce can be achieved.

We start with a spatially flat FRW metric
\begin{eqnarray}
 ds^2 = dt^2-a^2(t)d\vec{x}^2~,
\end{eqnarray}
where $a(t)$ stands for the scale factor. The dynamics of the FRW
universe is described by the Hubble parameter $H\equiv {\dot
 a}/{a}$ and its time derivative $\dot{H}$ which obey the well-known
Friedman equations:
\begin{eqnarray}
 H^2 &=& \frac{8\pi G}{3}\rho~,
\label{hubsq}\\
 \dot{H} &=& -4\pi G(\rho+p)~.
\label{doth}
\end{eqnarray}
At the moment of the non-singular bounce, i.e. at $t=t_B$, the Hubble
parameter vanishes $H(t_B)=0$ and in order to ensure that the universe
enters an expanding phase one also requires $\dot{H}(t_B)>0$ at the
same moment.

Let us consider that in the contracting phase the universe was
initially radiation dominated with the standard particles (with
$w=1/3$). As the scale factor contracts the temperature of the
Universe increases and the Lee-Wick particles (which are heavier than
their standard model partners) will gradually start to contribute
thermally as $t\rightarrow0^-$ and just before the bounce the energy
density of the universe would be given by Eq. (\ref{energyd2}). To
satisfy the first condition of bounce i.e. $H(t_B)=0$ one requires
$\rho(t_B)=0$ which can be seen from Eq.~(\ref{hubsq}). Thus, assuming
that just before the bounce all the particle species and their
Lee-Wick partners are thermalized, it can be seen from
Eq. (\ref{energyd2}) that at the time of bounce the energy density
vanishes i.e. $\rho(t_B)=0$ to yield
\begin{eqnarray}
\frac{\tilde{M}^2}{24}\tilde{g}_{*N}T_B^2 = \frac{7\pi^2}{240}
\tilde{g}_FT_B^4 +\frac{\pi^2}{30}nT_B^4~,
\end{eqnarray}
where $T_B$ is the temperature of the universe at bouncing time
$t_B$. This shows that at bounce the pressure given in
Eq. (\ref{pres2}) would be positive as
\begin{eqnarray}
p(t_B)=\frac{7\pi^2}{360}\tilde{g}_FT_B^4 +\frac{\pi^2}{45}nT_B^4\,>0~.
\end{eqnarray}
Thus at the bouncing point one has from Eq.~(\ref{doth}) that
\begin{eqnarray}
\dot H(t_B)=-4\pi G (\rho(t_B) + p(t_B))<0~,
\end{eqnarray}
which indicates that after the bounce the universe fails to enter into
an expanding phase. This means that in this simple cosmological model
which is comprised of pure radiation and their thermalized Lee-Wick
resonances, one cannot obtain a bouncing solution for a universe. Now
we will discuss two distinct cases which slightly deviate from this
simple scenario discussed above and where a bouncing universe scenario
can be realized.

\subsection{Lee-Wick infested radiation plasma and a CDM component}

First we consider a CDM component along with the radiation plasma
which is composed of thermal standard particles and their thermalized
Lee-Wick partners. The CDM candidate is non-relativistic throughout
the whole evolution and hence we can neglect its pressure in the
following analysis.  The CDM component too can have Lee-Wick partners
according to the Lee-Wick scenario, but these partners would be
heavier than the CDM particle under consideration. Hence the Lee-Wick
partners of the CDM component could never thermalize in the evolution
(as the CDM remain non-relativistic throughout) and would remain
off-shell. Thus these Lee-Wick partners of CDM component would not
contribute thermally to the energy density of the radiation
plasma. Thus the energy density and pressure of the CDM component
would be
\begin{eqnarray}
\rho_{\rm D} = n_{\rm D} m\,,\,\,\,\,p_{\rm D}\sim 0\,,
\label{dmeos}
\end{eqnarray}
respectively, where $n_{\rm D}$ and $m$ are the number density and
mass of the CDM particle respectively and $m \gg T$ during all the
phases of evolution of the universe.

\begin{figure}[h!]
\centering
\includegraphics[width=15cm,height=8cm]{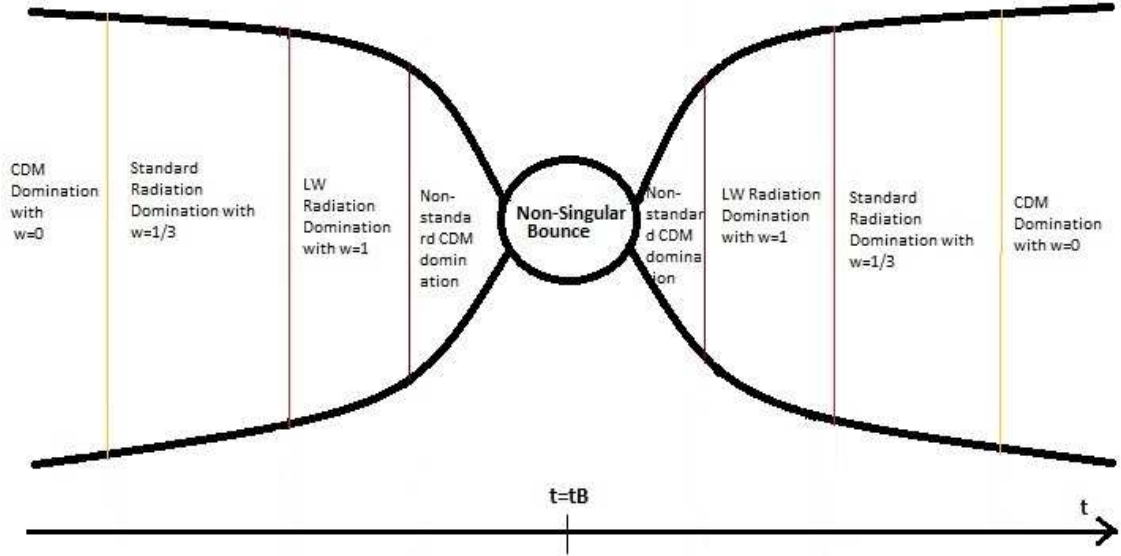}
\caption[]{Figure showing the bounce point and the energy balance of
  the universe. The CDM domination near the bounce point is due to the
  addition of the non-standard CDM like component in the theory. }
\label{bounce}
\end{figure}

Let us briefly illustrate the scenario under consideration. We naively
imagine an inverse picture of evolution of the expanding phase of the
universe in the contracting phase. Hence in the contracting phase the
universe initially evolves in a CDM dominated phase with $w=0$. As the
universe contracts the temperature of the universe increases and the
universe enters a standard radiation dominated phase with $w=1/3$. As
the temperature of the universe keeps on increasing with the
contraction the Lee-Wick resonances start to contribute thermally and
negatively to the energy density of the radiation bath. In that case,
though the temperature of the universe increases with contraction the
energy density starts to decrease with more and more Lee-Wick
particles (mainly the partners of chiral fermions and massless gauge
bosons) contributing thermally. The CDM component, during this period,
evolve non-relativistically with the background evolution decoupled
from the thermal bath. Once the energy density of the Lee-Wick
infested radiation bath becomes slightly smaller than the energy
density of the CDM component, the later dominates the energy density
of the universe and helps a bounce to occur, as will be shown
next. This is not a conventional matter domination phase as during
this period the state parameter of the cosmic fluid would not be
zero. This whole scenario is shown graphically in Fig. \ref{bounce}.

According to the above discussion, the total energy density of the
universe in this scenario, which consists of the energy density of the
Lee-Wick infested radiation plasma $\rho_R(t_B)$ and that of the CDM
particles $\rho_D(t_B)$, at the bouncing time $t_B$ turns out to be
\begin{eqnarray}
\rho(t_B)\equiv \rho_R(t_B)+\rho_D(t_B)=\frac{\tilde{M}^2}{24}
\tilde{g}_{*N}T_B^2 - \frac{7\pi^2}{240}
\tilde{g}_FT_B^4 - \frac{\pi^2}{30}nT_B^4 + \rho_{\rm D}(t_B)\,,
\label{energyd3}
\end{eqnarray}
whereas the expression for the pressure of this Lee-Wick infested
radiation plasma aided with the CDM particles remains the same as
given in Eq. (\ref{pres2}). If one wants to make $p(t_B)<0$, required
for a bounce at a temperature $T_{\rm B}$, then the first constraint
on the model is given by
\begin{eqnarray}
\frac{\tilde{M}^2}{24}\tilde{g}_{*N}T_{\rm B}^2 <
\frac{7\pi^2}{720}\tilde{g}_FT_{\rm B}^4 + \frac{\pi^2}{90}nT_{\rm B}^4\,.
\label{cons1}
\end{eqnarray}
If this constraint on the system is fulfilled, then one can indeed
have a bouncing universe where the energy density of the CDM candidate
has to satisfy the second constraint:
\begin{eqnarray}
\rho_{\rm D}(t_B) > \frac{7\pi^2}{360}\tilde{g}_FT_{\rm B}^4 +
\frac{\pi^2}{45}nT_{\rm B}^4\,,
\label{cons2}
\end{eqnarray}
which is obtained by requiring that $\rho(t_B)=0$ and using the first
constraint given in Eq.~(\ref{cons1}). Along with these two
constraints we also must have the other constraint about the
relativistic nature of the heaviest LW particle at $T_B$ which reads
as
\begin{eqnarray}
T_{\rm B} &> &\tilde{M},
\label{cons3}
\end{eqnarray}
This condition along with Eq.~(\ref{cons1}) yield the constraint on $\tilde{g}_{*N}$ as
\begin{eqnarray}
\tilde{g}_{*N}\geq \frac{7\pi^2}{30}\tilde{g}_F+\frac{4\pi^2}{15}n.
\end{eqnarray}

If we fix the free parameters of this model as $T_B\sim 10^{16}$ GeV
and roughly use the standard model values for calculating the number
of internal degrees of freedom as, $\tilde g_F\sim 40$ and $n\sim 10$,
then these three parameters yield $\rho_D(t_B)> 4\times 10^{64}$
GeV$^4$ and $\tilde{g}_{*N}\geq 100$.  Also to get $\rho(t_B)=0$ one
requires from Eq. (\ref{energyd3}) $\rho_R(t_B)=-\rho_D(t_B)$ which
yields $\tilde M\approx9\times 10^{15}$ GeV which is the mass of the
heaviest Lee-Wick particle in the model under consideration.

\subsection{Bouncing Universe within thermal LW framework without a 
CDM component}

We can have another scenario where only radiation can trigger the
bounce. But for this to happen some conditions have to be met. The
conditions are related to the mass of some Lee-Wick partners of the
normal particles and the bouncing temperature $T_B$. In this case we
consider that the bounce temperature $T_B$ is smaller than the masses
of some Lee-Wick particles. If this happens then some standard
particles' Lee-Wick partners will not be able to thermalize as their
masses are higher than $T_B$, but their standard partners being
lighter would be thermalized at $T_B$. Due to this effect these
(unpaired) standard particles will contribute only with positive
radiation energy density. Such standard particles whose Lee-Wick
partners could not thermalize at $T_B$ can both be bosonic and
fermionic. In this scenario the total energy density of the Universe
can be written as
\begin{eqnarray}\label{energyd4}
\rho(t_B)=\frac{\tilde{M}^2}{24}\tilde{g}_{*N}T_B^2 - \frac{7\pi^2}{240}
\tilde{g}_FT_B^4 - \frac{\pi^2}{30}nT_B^4 + \frac{\tilde c}{4} T_B^4\,,
\end{eqnarray}
and the total pressure would be
\begin{eqnarray}
p(t_B)=\frac{\tilde{M}^2}{24}\tilde{g}_{*N}T_B^2 -
\frac{7\pi^2}{720}\tilde{g}_FT_B^4 - \frac{\pi^2}{90}nT_B^4+ \frac{\tilde d}{4} T_B^4\,.
\label{pres4}
\end{eqnarray}
Here $\tilde c$ and $\tilde d$ encapsulate all the number of degrees
of freedom and the corresponding factors of those standard particles
whose Lee-Wick partners could not thermalize at bouncing point and
$\tilde M$ is the mass of the heaviest Lee-Wick particle which could
thermalize at or before $t_B$. Thus in this case too we have $\tilde
M< T_B$. To satisfy the bouncing condition $p(t_B)<0$ one requires
(using Eq. (\ref{pres4}))
\begin{eqnarray}
\frac{\tilde{M}^2}{24}\tilde{g}_{*N}T_B^2 <
\frac{7\pi^2}{720}\tilde{g}_FT_B^4 + \frac{\pi^2}{90}nT_B^4 - \frac{\tilde d}{4} T_B^4,
\label{cond1}
\end{eqnarray}
and to achieve the other bouncing condition i.e. $\rho(t_B)=0$ one gets
\begin{eqnarray}
 \tilde c-\tilde d> \frac{7\pi^2}{90}\tilde{g}_FT_B^4 + \frac{4\pi^2}{45}nT_B^4~,
\end{eqnarray}
where we have used Eq.~(\ref{energyd4}) and Eq.~(\ref{cond1}). This
scheme of a bouncing universe is also a possibility as for any
thermalized bosonic or fermionic standard model particle $p=\frac13
\rho$ and thus $\tilde d=\frac13 \tilde c$. Hence the above condition
can be satisfied by choosing an appropriate set of parameters.

\section{Cosmological perturbations in bouncing universe}

In the previous section two distinct thermal Lee-Wick scenarios are
considered where a non-singular bouncing universe scenario can be
achieved. We devote this section to study the dynamics of linear
cosmological perturbations generated in these models of thermal
Lee-Wick bounce and to investigate the nature of power spectrum
generated in such models. We focus on adiabatic fluctuations and
consider matter components without anisotropic stress (we refer to
\cite{Mukhanov:1990me} for a comprehensive review on the theory of
cosmological perturbations). In general, there are mainly two methods
of analyzing cosmological perturbations in bouncing scenarios. One is
to introduce a canonical variable of perturbation mode $v$ (known as
the Mukhanov-Sasaki variable) of which the quadratic action is of
canonical form in the frame of conformal time coordinate. This
variable is associated with the curvature fluctuation in comoving
gauge $\zeta$ through $v=z\zeta$ where $z$ is a background dependent
coefficient and is roughly proportional to the scale factor if the
background evolution is stable. A detailed calculation using a
generalized canonical variable $v$ in non-singular bounce cosmologies
was carried out in \cite{Cai:2012va} and it was shown that such a
method is better suitable for those early universe models in which the
primordial perturbations are originated from quantum
fluctuations. Another method is to study the gravitational potential
$\Phi$ directly. The advantage of this method over the other is that
it is much easier to impose initial conditions while analyzing
gravitational potential $\Phi$ as cosmological perturbations
\cite{Cai:2009rd}. We will follow the second method for our analysis
of cosmological perturbations.
\begin{figure}[h!]
\centering
\includegraphics[width=12cm, height=7cm]{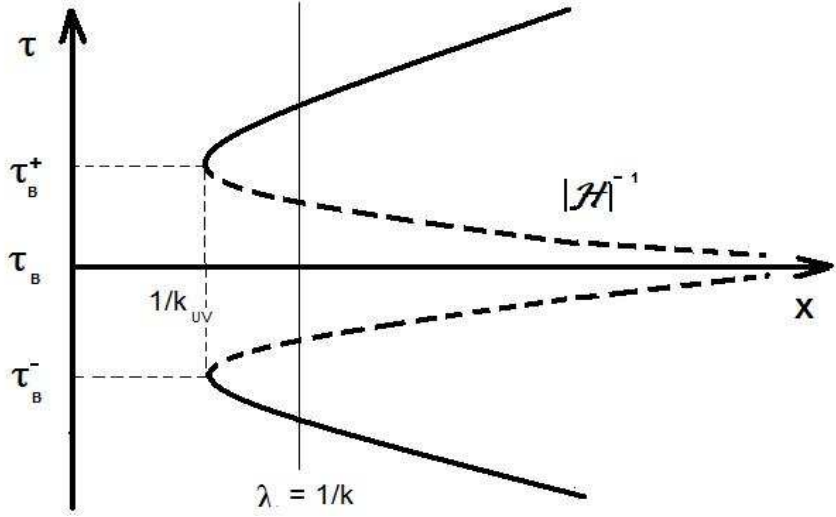}
\caption[]{Figure showing the contracting and expanding phases of the
  universe in the bouncing model. The horizontal axis is a comoving
  spatial coordinate and the vertical axis is conformal time. The main
  plot is of the Hubble radius $|{\cal H}|^{-1}$ and the wavelength
  $\lambda$ of fluctuations with comoving wave number $k$. $k_{\rm
    UV}$ is roughly the wave number at the onset of the bouncing phase
  whose value is dictated by the microphysics of the bounce. The heavy
  dashed curves on the main figure corresponds to the bounce phase. }
\label{bounce-hubble}
\end{figure}

Before going into the details of the analysis of cosmological
perturbations and the corresponding power spectra let us discuss
briefly how the modes of these primordial perturbations evolve in the
contracting and the expanding phases. For the discussion let us only
consider the case with CDM component and Fig.~\ref{bounce-hubble}
explains the essence of this particular bouncing model. Similar
analysis also holds for the other case which does not consider any CDM
component. In Fig.~\ref{bounce-hubble} it is shown how the Hubble
radius $|{\cal H}|^{-1}$ behaves in the space-time diagram. The lower
most heavy curve shows that the universe is undergoing a contracting
phase where the energy density of the universe is changing from the
standard radiation dominated phase ($\rho \propto T^4$) to the mildly
Lee-Wick radiation phase ($\rho \propto T^2$). In this phase, the dark
matter energy density is not important as it is mainly a radiation
dominated phase. At the beginning of this Lee-Wick infested radiation
phase, as all the Lee-Wick partners are not thermalized, the effective
energy density will look like
\begin{eqnarray}
\rho \sim  aT^4 + bT^2 - c T^4\,,
\label{effrho}
\end{eqnarray}
where $a$, $b$ and $c$ are some constants whose exact value is not of
much importance in this discussion. The first term $aT^4$ is the
dominant term in the standard radiation domination as shown in
Fig. \ref{bounce} and arises because the radiation domination is
mainly dictated by the standard fields. The $bT^2$ term designates
that some of the Lee-Wick partners have entered the scene and they
have paired up with their normal partners to give the $T^2$ nature of
energy density. The last term $cT^4$ is the contribution from the
unpaired Lee-Wick partners of chiral fermions and massless gauge
bosons. The schematic form of energy density, in Eq. (\ref{effrho}),
is similar to the form of energy density in Eq. (\ref{energyd2})
except that here one also takes into account the standard thermal
radiation part ($aT^4$). In presenting the energy density of Lee-Wick
thermal radiation in Eq. (\ref{energyd2}) one assumed that all of the
standard field's Lee-Wick partners are in thermal equilibrium. In the
initial stage of the contracting phase of the universe when the
temperature of the system is not very high it is natural that most of
the Lee-Wick partners would not have thermalized and so
Eq. (\ref{effrho}) holds true. In this phase $b \gg c$ and $a \gg c$,
which amounts to an assumption that most of the Lee-Wick partners of
the chiral fermions and the massless gauge bosons (which give rise to
the negative radiation energy density) are predominantly heavier than
the other Lee-Wick partners. If this was not the case then the
Lee-Wick partners of the chiral fermions and massless gauge bosons
should have dominated the energy-density at the onset of Lee-Wick
infested radiation era and the total energy density of the universe
should have decreased dramatically.

At the start of the contracting phase, when $a>b$, the state parameter
$w \sim 1/3$ designating a standard radiation phase. After some time
$b>a$ (and $b\gg c$) and the state parameter becomes $w \sim 1$. This
phase is shown as the Lee-Wick radiation domination phase in
Fig. \ref{bounce}.  As because of the $b$ and $c$ terms in the energy
density the effective energy density of this phase rises but its rise
is slowed down (as a result the growth of the Hubble parameter is
inhibited) and consequently the Hubble radius goes on decreasing but
with a slower rate. During this phase the modes with wavelength
$\lambda=1/k$ leaves the Hubble radius as shown in the
Fig. \ref{bounce-hubble}.

As soon the conformal time is about $\tau_B^-$ the $c$ term in the
effective energy density starts to dominate and now $c > a\,,\,b$ and
the radiation part of the effective energy density starts to decrease
rapidly. In this phase the radiation energy density starts to decrease
as more and more Lee-Wick partners of the chiral fermions and massless
gauge bosons thermalize and the Hubble parameter effectively decreases
rapidly and the CDM energy density starts to dominate. This phase of
the universe is represented by the non-standard CDM domination phase
in Fig. \ref{bounce}. The CDM contribution to the energy density
ultimately balances the negative energy density due to Lee-Wick
radiation and leads to a bounce.

If we assume symmetry in time then a symmetric analysis can be
presented for the expanding phase of the universe. The heavy dashed
curves from $\tau_B^-$ to $\tau_B^+$ corresponds to the bouncing phase
of the universe where the microphysics is mainly governed by the
Lee-Wick partners of chiral fermions and massless gauge bosons and the
dark matter sector. The solid heavy curves stand for the contracting
and expanding phases of the universe. The modes with wavelength
$\lambda$ which went out during the contracting phase starts to
re-enter the Hubble radius after $\tau_B^+$ in the expanding phase. In
Fig. \ref{bounce-hubble}, $\tau_B$ is the time of bounce where as
$\tau_B^\pm$ stands for the time where there should have been a Big
Bang or Big Crunch like singularities in absence of the bounce. In
Fig. \ref{bounce-hubble}, $k_{\rm UV}$ is the maximum possible wave
number, which becomes super-Hubble, in such a bouncing scenario as it
corresponds to the minimum radius of the possible Hubble radius
possible. In absence of the bounce $k_{\rm UV}$ could have been much
bigger, effectively of the Planckian order.

\subsection{Cosmological perturbations in contracting and expanding phases 
and their matching conditions}

Here we will discuss how cosmological perturbations generate in a
contracting phase and evolve into the expanding phase in a generic
bouncing universe scenario. The linearly perturbed FRW metric in
longitudinal gauge can be written as \cite{Mukhanov:1990me}
\begin{eqnarray}
 ds^2 = a(\tau)^2 \left[(1+2\Phi)d\tau^2 - (1-2\Phi)d\vec{x}^2\right]~,
\end{eqnarray}
where $\Phi$ is the gravitational potential which characterizes metric
fluctuations and $\tau$ is the conformal time. At linear order the
scalar metric fluctuations evolve independently. Thus we are able to
study the evolution of $\Phi$ by following one of its Fourier mode
with a fixed comoving wave number $k$ and the perturbation equation of
$\Phi_k$ is given by
\begin{eqnarray}
 \Phi_k''+2\eta{\cal H}\Phi_k'+\left(c_s^2k^2 -2\epsilon{\cal H}^2 +
2\eta{\cal H}^2\right)\Phi_k = 0~,
\label{pert1}
\end{eqnarray}
where ${\cal H}\equiv a'/a$ and the prime denotes the derivative with
respective to $\tau$. The sound speed parameter $c_s$ in the above
equation is usually determined by thermodynamical property of the
background system. Moreover, we have defined two background dependent
parameters $\epsilon\equiv -\dot{H}/H^2$ and
$\eta\equiv-\ddot{H}/2H\dot{H}$. For a constant background equation of
state (say $w$) these two parameters are equal and are totally
determined by the background equation of state as
\begin{eqnarray}
\epsilon=\eta=\frac32(1+w)~.
\label{pert2}
\end{eqnarray}

The situation which we are interested in here is a contracting phase
of the universe dominated by radiation fields along with their
thermalized Lee-Wick partners evolving in a stable background which
evolves with a constant state parameter $w$. In that case the scale
factor and the Hubble parameter of the Universe can be written as
\begin{eqnarray}
 a\sim (\tau-\tau_B^i)^{\frac{2}{1+3w}}~,~~{\cal H} \simeq \frac{2}{(1+3w)
(\tau-\tau_B^i)}
\label{a-and-H}
\end{eqnarray}
respectively, where $\tau_B^i$ is some moment at which the Big Bang or
the Big Crunch singularity would occur in absence of the non-singular
bouncing phase. In our particular model $\tau_B^i =
\tau_B^+\,,\,\tau_B^-$ as shown in Fig. \ref{bounce-hubble}. As a
consequence, the perturbation equation given in Eq. (\ref{pert1}) can
be simplified using Eq. (\ref{pert2}) as
\begin{eqnarray}\label{pertEoM}
 \Phi_k''+\frac{1+\nu}{\tau-\tau_B^i}\Phi_k' +c_s^2k^2\Phi_k \simeq 0~,
\end{eqnarray}
with
\begin{eqnarray}\label{nu}
 \nu = \frac{5+3w}{1+3w}~.
\end{eqnarray}
Generically, the solutions to Eq.~\ref{pertEoM} are composed of two
linearly independent Bessel functions. On super-Hubble scales they
correspond to a constant mode (called as the D mode) and a
time-evolving mode (called as the S mode). Specifically, the general
solution to the perturbation equation on super-Hubble scales can be
expressed as
\begin{eqnarray}
 \Phi_k^{\pm} = D_{\pm}+S_{\pm}\left(\frac{\tau_B-\tau_B^{\pm}}
{\tau-\tau_B^{\pm}}\right)^{\nu}~,
\label{Phi-pm}
\end{eqnarray}
where $D$ and $S$ are the mode coefficients and the subscripts $\pm$
represent the expanding and the contracting phase of the universe
respectively. In the above expression $\tau_B$ is the conformal time
at the bouncing point. It is also assumed in the above solutions that
the background dynamics of the universe both in the contracting and in
the expanding phases are governed by the same steady equation of state
$w$.

Now one needs to know how to transfer the primordial fluctuations
generated in the contracting phase of the universe through the
bounce. This issue was initially studied by replacing the bounce with
a matching surface across which the perturbation modes are connected
by using the Hwang-Vishniac \cite{Hwang:1991an} (or Deruelle-Mukhanov
\cite{Deruelle:1995kd}) matching conditions. Later, it was found that
for a non-singular bounce one can evolve the fluctuations through
bounce both numerically and analytically \cite{Cai:2008qw, Cai:2007zv,
  Cai:2008ed}. Thus, relations between the mode coefficients in the
expanding phase and those in the contracting phase can be calculated
explicitly \cite{Cai:2008qw} (also see \cite{Cai:2009rd} for a general
discussion\footnote{It is interesting to observe that for a
    nonsingular bounce model based on a closed space geometry, the
    primordial perturbation would be dramatically affected by the
    spatial curvature during the bounce even at large scale
    limit\cite{Martin:2003sf}. However, since in the model under
    consider the bouncing solution is achieved by introducing matter
    components which violate null energy condition in the frame of
    flat spatial coordinates, the curvature term will not be involved
    in the following transfer relation, as has been verified
    explicitly in the cold Lee Wick bounce model
    \cite{Cai:2008qw}.}). A general transfer relation between these coefficients in
the contracting and expanding phase can be written as
\begin{eqnarray}\label{matching}
 D_+ = {\cal O}(1) D_- + {\cal O}(1) \left(\frac{k}{k_{\rm UV}}\right)^2 S_-~,
\end{eqnarray}
where $k_{\rm UV}$ is a normalization scale which is set by the
micro-physics of the bounce. It corresponds to the inverse length
scales near around $\tau_B^\pm$ in our model as shown in
Fig.~\ref{bounce-hubble}. A brief derivation of the above relation is
presented in Appendix~\ref{transfer}.

\subsection{Thermal fluctuations in the case which includes CDM}

In the model which we consider here the universe experiences a
Lee-Wick infested radiation dominated period in contracting phase just
before the CDM component takes over to yield a bounce. In the
following we will study the generation of primordial power spectrum of
curvature perturbation arisen from thermal fluctuations of Lee-Wick
infested radiation plasma. Due to the existence of Lee-Wick partners,
the heat capacity of the radiation plasma is different from the
conventional one and thus affect the scale dependence of the power
spectrum. Then we will study the stability of this cosmological system
and estimate the mass bounds of the Lee-Wick partners.

\subsubsection{Generation of primordial power spectrum of thermal 
fluctuations in IR regime}

To determine the IR regime of the primordial power spectrum of the
thermal fluctuations generated in the contracting phase of the
universe, one should consider the modes which leave the horizon at
earliest times of the contracting phase. From Fig.~\ref{bounce}
we see that at the onset of radiation dominated era in the contracting
phase the radiation fluid is dominated mainly by standard particles
(as Lee-Wick partners being heavy thermalize at later stage) yielding
$w_r\approx1/3$. The radiation energy density during that time
thus can be written as
\begin{eqnarray}
\rho_r\approx g_{\rm sm}T^4/4,
\label{sm-rad}
\end{eqnarray}
where $g_{\rm sm}$ takes into account the number of degrees of freedom
of the standard particles thermalized initially. We also consider that
the background during these initial phases of radiation domination
still evolves under the influence of the CDM yielding $w\sim 0$ and
the index $\nu=5$ introduced in Eq.~(\ref{nu}).

The correlation function of the energy density of such a cosmic fluid
in thermal equilibrium can be written as
\begin{eqnarray}\label{ic_thermal}
 \langle \delta\rho^2 \rangle |_{R} =\frac{k^3}{2\pi^2}\langle\delta
\rho_k^2\rangle = C_V \frac{T^2}{V^2} ~,
\end{eqnarray}
where $V$ is a fixed volume determined by the correlation length $R$
of the thermal system which in a cosmological setup is roughly of the
same order of the Hubble radius $R\sim 1/H$ and a Fourier transform of
any generic quantity $\phi$ with the corresponding Fourier mode
$\phi_k$ is defined as
\begin{eqnarray}
\phi(t,{\mathbf x})=\sqrt{V}\int\frac{d^3{\mathbf k}}{(2\pi)^{3/2}}
e^{i\mathbf{k\cdot x}}\phi_k(t).
\end{eqnarray}
The parameter $C_V$ is the heat capacity of the radiation plasma and
is defined in terms of the expectation value of the internal energy
within the thermally correlated volume. These fluctuations in the
energy density of the radiation fluid are coupled to the metric
perturbations $\Phi_-$ during the contracting phase through the
time-time component of the perturbed Einstein equation which is given
as \cite{Mukhanov:1990me}
\begin{eqnarray}\label{pert00}
 -3{\cal H}({\cal H}\Phi_-+\Phi_-')+\nabla^2\Phi_- = a^2\delta\rho/2M_{p}^2~,
\end{eqnarray}
where $M_P$ is the reduced Planck mass. When these perturbations leave
the Hubble radius during the contracting phase (i.e. when the
corresponding wavenumber of the mode is of the order of the comoving
Hubble radius $k\sim a H$), the three terms contribute equally to the
amplitude of $\Phi_-$ yielding
\begin{eqnarray}
|\Phi_k^-|\sim \frac{a(\tau_k)^2\delta\rho_k}{2k^2M_{p}^2}\simeq \frac{\pi a^2(\tau_k)C_V^{1/2}(\tau_k) T(\tau_k)}{2^{1/2}k^{7/2}M_p^2V(\tau_k)} ~,
\label{ic_metric}
\end{eqnarray}
up to a constant of order $\mathcal{O}(1)$. We have used
Eq.~(\ref{pert00}) to derive the second expression and $\tau_k$ is the
conformal time at the time of Hubble-crossing of the mode $k$. All the
quantities in the above equation is to be derived at the time of
Hubble crossing of the mode $k$.

Now, the key issue is to find out the explicit form of heat capacity
of the thermal system under consideration. By definition, the heat
capacity of a thermal fluid is determined by
\begin{eqnarray}
 C_V \equiv V\frac{\partial\rho}{\partial T} ~.
\label{cv-general}
\end{eqnarray}
As has been mentioned before, the volume $V$ is determined by the
thermal correlation length which is in order of the Hubble
radius. Consequently, using Eq.~(\ref{sm-rad}) the heat capacity at
the Hubble-crossing moment of the IR modes is given by
\begin{eqnarray}
 C_V (\tau_k) \approx \frac{4\pi}{3} g_{\rm sm}(\tau_k)\frac{T(\tau_k)^3}{H(\tau_k)^3}~.
\end{eqnarray}
We note here that the perturbations in the radiation energy density
scales with the scale factor $w_r$ as
\begin{eqnarray}
\delta\rho\sim a^{-3(1+w_r)},
\end{eqnarray}
where the entropy density of a fluid is defined as
\begin{eqnarray}
s=\frac{\delta\rho+\delta p}{T}=(1+w_r)\frac{\delta\rho}{T}\,.
\end{eqnarray}
For an adiabatic expansion of the Universe the comoving entropy
remains conserved which means $a^3 s$ is a conserved quantity. Thus
the temperature of the radiation plasma evolves as
\begin{eqnarray}
T\sim a^{-3w_r},
\end{eqnarray}
which follows from the previous two equations.  Moreover, the scale
factor of the universe evolves as $a\sim\tau^{\frac{1}{2}(\nu-1)}$ in
the contracting background which can be seen from Eq. (\ref{a-and-H})
and at the time of Hubble crossing of the mode $k$ one has
$\tau_k=-\frac1k$. Then we have $a=a_*\left(k/k_*\right)^{-2}$.
Thus, following the above expressions, the temperature of the universe
at the time when the mode $k$ leaves the Hubble radius can be
determined as
\begin{eqnarray}\label{T_k}
 T(\tau_k) \simeq T_* \left(\frac{k}{k_*}\right)^{\frac{3w_r}{2}(\nu-1)}~,
\end{eqnarray}
where $k_*$ and $T_*$ are associated with the initial moment of
thermal equilibrium as introduced in previous subsection. Similarly
the Hubble parameter $H(\tau_k) \equiv
\frac{\mathcal{H}(\tau_k)}{a(\tau_k)}$ during that time can also be
derived using Eq.~(\ref{a-and-H}) which reads as
\begin{eqnarray}\label{H_k}
H(\tau_k) \simeq H_* \left(\frac{k}{k_*}\right)^{\frac{1}{2}(\nu+1)}~,
\end{eqnarray}
where $H_*$ again is associated with $k_*$. Thus the above two
equations yield
\begin{eqnarray}
 T(\tau_k) \simeq T_* \left(\frac{k}{k_*}\right)^2~, ~~H(\tau_k) \simeq H_*\left(\frac{k}{k_*}\right)^3~,
\end{eqnarray}
if the perturbation modes exit the Hubble radius before the universe
is fully thermalized. Here, $k_*$ is a normalization scale which is
associated with the initial moment of thermal equilibrium, $T_*$ is
the corresponding temperature at that moment, and $H_*$ is the Hubble
parameter corresponding to $k_*$ as well. Thus the heat capacity is
given by
\begin{eqnarray}
 C_V(\tau_k) \approx \frac{4\pi}{3} g_{\rm sm}(\tau_k) \frac{T_*^3}{H_*^3} \left(\frac{k}{k_*}\right)^{-3} ~,
\end{eqnarray}
at the moment of Hubble crossing during initial radiation
contraction. Then we get the amplitude of gravitational potential as
\begin{eqnarray}\label{ic_model0}
 \Phi_H \equiv \Phi_k^-(\tau_H) \simeq
 \frac{(3\pi)^{1/2}}{2^{3/2}M_p^2} 
\frac{\sqrt{g_{\rm sm}}\,\,T_*^{5/2}}{a_*^{7/2}H_*^{4}}k^2~,
\end{eqnarray}
where Eq. \eqref{ic_metric} was applied and the scale factor $a_*$ is
associated with the moment of thermal equilibrium. Following
Eq.~(\ref{Phi-pm}), we see that the amplitudes of the constant
$D_-$ mode and the growing $S_-$ mode during the initial radiation
contraction are related to the metric fluctuation as
\begin{eqnarray}
 D_-(k) \simeq \Phi_H~,~~S_-(k) \simeq
\left(\frac{{\cal H}_{\rm UV}}{k}\right)^5 \Phi_H~.
\end{eqnarray}
Here ${\cal H}_{\rm UV}$ is the maximal scale of the Hubble parameter,
in the beginning of the bouncing phase, which is roughly of the same
order of $k_{\rm UV}$ introduced in Eq.~(\ref{matching}). Thus from
the above equation and Eq.~(\ref{matching}) we easily find out that
the amplitude of the $D_+$ mode in expanding phase is mainly
contributed by the $S_-$ mode in IR regime. Correspondingly, we derive
the primordial power spectrum of metric perturbation in IR regime as
\begin{eqnarray}\label{P_Phi_0}
 P_\Phi = \frac{k^3}{2\pi^2} |D_+|^2
 \simeq \frac{3 g_{\rm sm}\mathcal{H}_{\rm UV}^6 T_*^5}{16\pi M_p^4 a_*^6 H_*^7} \left(\frac{k}{k_*}\right)~,
\end{eqnarray}
and the corresponding spectral index is given by
\begin{eqnarray}\label{n_Phi_0}
 n_\Phi \equiv 1+\frac{d\ln P_\Phi}{d\ln k} = 2~,
\end{eqnarray}
which implies that the power spectrum of metric perturbations seeded
by thermal fluctuations during the onset of radiation dominated
contracting phase is a blue spectrum.

\subsubsection{Generation of primordial power spectrum of thermal 
fluctuations in UV regime}

As we have pointed out before, the thermodynamics of the radiation
fluid in the contracting phase would be greatly affected by the
Lee-Wick resonances at very high temperatures ($T>\tilde M$) and the
corresponding equation of state of the radiation fluid evolves from
the conventional value of $w_r=1/3$ to a non-conventional value of
$w_r\approx1$ when more and more Lee-Wick resonances thermalize with
the increasing temperature of the universe. Therefore, the generation
of primordial power spectrum of thermal fluctuations is dramatically
changed in the UV regime as the background and the radiation fluid at
these later stage of the evolution will evolve with state parameters
as $w \sim w_r\sim1$.

During the Hubble-crossing of the mode $k$ the cosmic fluid would be
dominated by Lee-Wick resonance infested radiation plasma and the
energy density of that fluid can be expressed as given in
Eq. (\ref{energyd2}) (where we can neglect the contribution of the CDM
component required for the non-singular bounce). Also, to let the
Hubble radius of the universe contract which allows the mode to become
super-Hubble, the first term in Eq. (\ref{energyd2}) would be the
dominant term in the total energy density (as discussed before). If
during that time the mass of the heaviest thermalized Lee-Wick
particle is $\tilde M$ then using Eq. (\ref{energyd2}) in
Eq.~(\ref{cv-general}) one gets an expression for the heat-capacity of
the radiation fluid at the time of Hubble-crossing of the mode $k$ as
\begin{eqnarray}\label{heat}
 C_V(\tau_k) \approx V(\tau_k) \alpha(\tau_k) \tilde M^2 T(\tau_k) ~,
\end{eqnarray}
where we have introduced
\begin{eqnarray}
 \alpha(\tau_k) = \frac{\tilde{g}_{*N}(\tau_k)}{12}~,
\end{eqnarray}
which is associated with the number of internal degrees of freedom
of the thermal system at the time $\tau_k$.  As the energy density and
pressure of the radiation fluid is dominated by the $\alpha$ term (as
given in Eq.~(\ref{energyd2})) during the Hubble-crossing of the modes
the equation of state of that plasma would be
$w_r(\tau_H)\approx1$. In that case, following Eq. (\ref{heat}), the
heat capacity of the universe would be
\begin{eqnarray}
 C_V(\tau_k)\approx \frac{4\pi}{3}\alpha(\tau_k)\tilde M^2 \frac{T_*}{H_*^3}\left(\frac{k}{k_*}\right)^{-3},
\end{eqnarray}
where we have used Eq.~(\ref{T_k}) and Eq.~(\ref{H_k}). Thus, if we
also consider in this scenario that the background cosmology of the
universe is also evolving due to the dominance of the Lee-Wick
infested radiation field and consider $w=w_r\approx1$ during the
Hubble-exit of the mode $k$, then we get the amplitude of the metric
fluctuation corresponding to mode $k$ using Eq.~(\ref{ic_metric}) as
\begin{eqnarray}\label{ic_model1}
 \Phi_H \equiv \Phi_k^-(\tau_H) \simeq \frac{(3\pi)^{1/2}}{2^{3/2}M_p^2}
\frac{\sqrt{\alpha}\tilde M T_*^{3/2}}{a_*^{3/2}H_*^{2}}~,
\end{eqnarray}
where $a_*$ is the scale factor corresponding to $k_*$ through
$k_*=a_*H_*$. Following Eq.~(\ref{ic_model1}), we see that the
amplitudes of the constant $D_-$ mode and the growing $S_-$ mode
during the contracting phase are related to the metric fluctuation as
\begin{eqnarray}\label{DS-}
 D_-(k) \simeq \Phi_H~,~~S_-(k) \simeq
\left(\frac{{\cal H}_{\rm UV}}{k}\right)^2 \Phi_H~,
\end{eqnarray}
where $\nu=2$ as $w=1$ in this particular case. Here ${\cal H}_{\rm
  UV}$ is the maximal scale of the Hubble parameter, in the beginning
of the bouncing phase, which is roughly of the same order of $k_{\rm
  UV}$ introduced in Eq.~(\ref{matching}). Thus from the above
equation and Eq. (\ref{matching}) we see that both the modes $D_-$ and
$S_-$ contribute equally to the amplitude of the $D_+$ mode which is
the dominant mode of the expanding phase (as the other $S_+$ mode is
the decaying one and its amplitude falls of with time)\footnote{We
  would like to refer to Refs. \cite{Cai:2008qw, Cai:2012va,
    Cai:2007zv, Cai:2008ed, Cai:2011tc} for extensive analysis on
  microscopic description of nonsingular bouncing phase in a wide
  class of linear bounce models.}. Hence, we obtain the primordial
power spectrum of the expanding phase of the universe as
\begin{eqnarray}
 P_\Phi = \frac{k^3}{2\pi^2} |D_+|^2
 \simeq \frac{\tilde{g}_{*N} \tilde M^2 T_*^3}{64\pi M_p^4 H_*}
\left(\frac{k}{k_*}\right)^3~,
\end{eqnarray}
and the corresponding spectral index is given by
\begin{eqnarray}
 n_\Phi \equiv 1+\frac{d\ln P_\Phi}{d\ln k} = 4~,
\end{eqnarray}
which implies that the power spectrum of metric perturbations seeded
by thermal fluctuations during the Lee-Wick resonance dominated
contracting phase is highly blue in UV regime.

\subsubsection{Stability analysis and a rough estimate of the mass bounds of the Lee-Wick partners of chiral fermions and massless gauge bosons}

As the primordial power spectrum of metric perturbation generated in
Lee Wick radiation phase is deeply blue, its amplitude would become
secondary in infrared limit which corresponds to large length
scale. However, one needs to be aware of the potential concern that
the power spectrum would become too large in ultraviolet regime where
$k$ takes a large value.  Fortunately, for all nonsingular bounce
models, there is a natural ultraviolet cutoff on $k$ modes due to the
existence of the bouncing scale. This is because, if a perturbation
mode has not yet evolved into super-Hubble scale before the bouncing
phase, then it will never take place unless there is a period of
inflation after the bounce. As a consequence, one can easily read that
the maximal value of the power spectrum takes place right at the
beginning of the bouncing phase.

The universe is in thermal equilibrium, with standard particles, in
the phase of traditional radiation domination and thus one can assume
the benchmark value of the Hubble parameter to be, $H_*\simeq
\frac{g^{1/2}\pi T_*^2}{3\sqrt{10}M_p}$ where $g$ is the number of
internal degree of freedom of traditional radiation. Using this
equation and the relation between $H$ and $k$ from Eq.~(\ref{H_k}),
one can calculate the maximal value of the primordial power spectrum
as:
\begin{eqnarray}
 P_\Phi^{\rm max} \simeq \frac{27 \tilde{g}_{*N} {\tilde M}^2 H_{\rm UV}^2}
{64\pi g^{3/2} M_p T_*^3}~.
\end{eqnarray}
In the above equation $H_{\rm UV}$ is roughly similar to $k_{\rm UV}$
whose value is set by the microphysics of the bounce. In our notation
it is trivial to show that
\begin{eqnarray}
 \langle\Phi({\bf x})^2\rangle = V \int_0^{k_{\rm UV}} \frac{dk}{k}P_\Phi(k)~,
\end{eqnarray}
where $P_\Phi(k)={k^3}|\Phi_k|^2/2\pi^2 $ is the power spectrum of
metric perturbation $\Phi$. Note that, this UV cutoff can be selected
as the Planck scale $k_{\rm UV} \sim a M_{p}$ in inflationary
cosmology, and due to the quasi exponential expansion of the
background universe, it leads to the well-known Trans-Planckian
problem for inflationary perturbation \cite{Martin:2000xs,
  Brandenberger:2000wr} (see also \cite{Brandenberger:1999sw}). This
issue does not happen in this model of bouncing cosmology as there is
another natural cutoff, $k_{\rm UV}$ as shown in
Fig. \ref{bounce-hubble} which is much lower than Planck scale, i.e.,
the bounce scale (the maximal absolute value that the Hubble parameter
can reach throughout the whole evolution). If $P_\Phi(k)\propto k^n$
(where $n=3$ in our bounce model) then we obtain, $\langle\Phi({\bf
  x})^2\rangle \sim \frac{k_{\rm UV}^n}{n} \sim \frac{P_\Phi^{\rm
    max}}{n} ~.$ As $\Phi({\bf x})$ is supposed to be metric
perturbation we expect $\frac{1}{V} \langle\Phi({\bf x})^2\rangle$ to
be smaller than unity. Consequently the above equation implies the
maximal value of the power spectrum to be less than unity as well.

If we take $g\simeq 107$ which is the number of internal degree of
freedom of standard model of particle physics, then we can roughly
obtain an upper bound on a combination of the Lee Wick mass and the
bounce scale as
\begin{eqnarray}
{\tilde M} |H_{\rm UV}| \lesssim 100 \left(\frac{M_p T_*^3}
{\tilde{g}_{*N}}\right)^{1/2}~,
\end{eqnarray}
by requiring $P_\Phi^{\rm max} \lesssim 1$. We further take
$\tilde{g}_{*N}\sim 100$ and assume $T_*\sim 100~ {\rm GeV}$, and then
get ${\tilde M} |H_{\rm UV}| \lesssim {10^{14}~ {\rm
    GeV}^2}$. However, since the bounce scale must be higher than the
mass scale of Lee Wick partners $|H_{\rm UV}|>\tilde{M}$, it roughly
implies that $\tilde{M} \lesssim 10^7 ~ {\rm GeV}$.

This theoretical constraint restricts the masses of the Lee-Wick
partners in the expanding or in the contracting phase of the universe
to be less than $10^7 ~ {\rm GeV}$ as only the terms proportional to
$\alpha$ is taken into account in the heat capacity. The constraint
does not work in the bouncing phase of the universe (in the time
between $\tau_B^-$ and $\tau_B^+$) where much heavy Lee-Wick partners
can get thermalized. In the present models where a dark component is
used we have calculated the maximum mass of the Lee-Wick partner to be
of the order of $10^{15} ~ {\rm GeV}$, but this does not contradict
the analysis given above as the heavily massive Lee-Wick particle is
only thermalized at the bounce point, near $\tau_B$. Predominantly in
the bouncing phase the Lee-Wick partners of the chiral fermions and
mass less gauge bosons will contribute and consequently we can infer
that most of the Lee-Wick partners of the chiral fermions and mass
less gauge bosons will have their mass in the range $10^7 ~ {\rm
  GeV}\, -\, 10^{15} ~ {\rm GeV}$ if the temperature near the bounce
is $10^{16} ~ {\rm GeV}$.

\subsubsection{Numerical analysis of the background and the
  perturbation in the bouncing phase} 

Previously the problem of mode matching near the bounce was done
analytically for modes in the IR or UV regime. In the analytic way one
matches the modes which were super Hubble before $t_B^-$ and after
$t_B^+$. The super Hubble modes briefly re-enter the causal horizon
during the bouncing phase (in the time interval between $t_B^+$ and
$t_B^-$). To see what happens inside the bouncing phase we use
numerical techniques.
\begin{figure}[ht!]
\centering
\includegraphics[width=10cm, height=7cm]{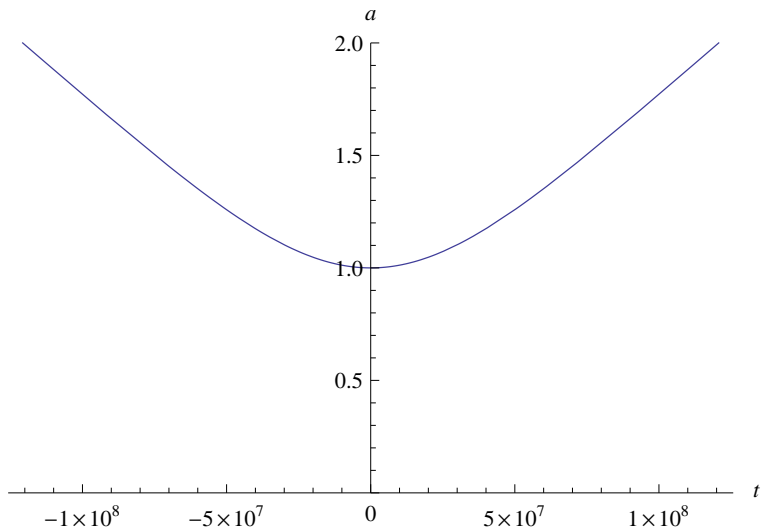}
\caption[]{Figure showing the behavior of the scale factor $a(t)$ in
  the bouncing phase.
}
\label{atp}
\end{figure}

To ensure that the model under consideration indeed yields a bouncing
phase of the Universe we analyze the behavior of the background
plasma on which the perturbations live. It is seen that with some
benchmark values of the theory the model gives rise to predictable
bouncing behavior.
\begin{figure}[ht!]
\centering
\includegraphics[width=10cm, height=7cm]{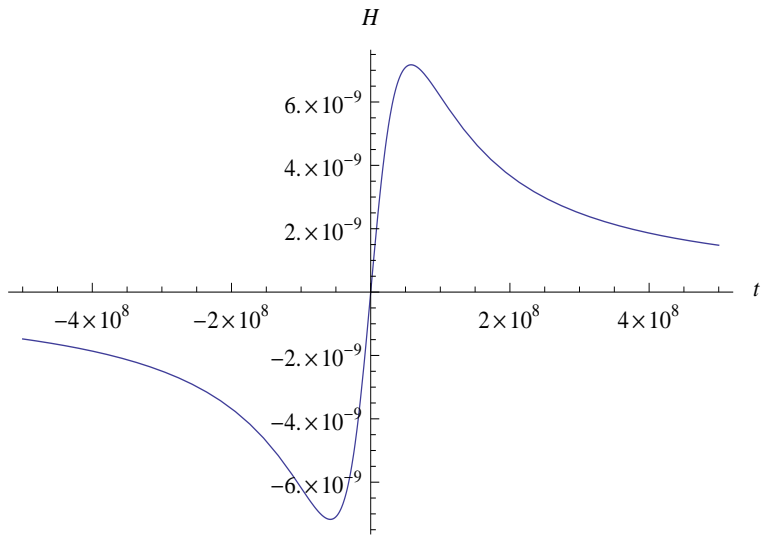}
\caption[]{Figure showing the behavior of the Hubble parameter $H(t)$ in
  the bouncing phase.
}
\label{Htp}
\end{figure}
In the numerical code we set the scale factor $a(t)=1$ at the bouncing
time $t=t_B \equiv 0$. We also set the temperature of bounce, $T_B$,
and the mass of the heaviest Lee-Wick partner's mass during the
bouncing phase, ${\tilde M}$, as
\begin{eqnarray} 
{\tilde M} = 10^{-5}\,M_p\,,\,\,\,T_B=10^{-4}\,M_p\,,
\label{mtb}
\end{eqnarray}
where $M_p$ is the Planck mass which is set to one, $M_p=1$. In this
numerical analysis the time unit is set to be the Planck time.
The other parameters appearing in Eq.~(\ref{energyd3}) are set as
\begin{eqnarray} 
\tilde{g}_{*N} = 120\,,\,\,\,\tilde{g}_{F}=40\,,\,\,\,n=10\,.
\label{degf}
\end{eqnarray}
It is to be noted that as we treat the CDM component to be
non-relativistic throughout the evolution of the universe, the energy
density of this CDM component evolves as
\begin{eqnarray}
\rho_D(t)=\frac{\rho_B}{a(t)^3}\,,
\label{rhodb}
\end{eqnarray}
where the constant $\rho_B$ is found from the condition that at the
bouncing time $t=t_B$ one must have $\rho_R(t_B)+\rho_D(t_B)=0$.
\begin{figure}[ht!]
\centering
\includegraphics[width=10cm, height=7cm]{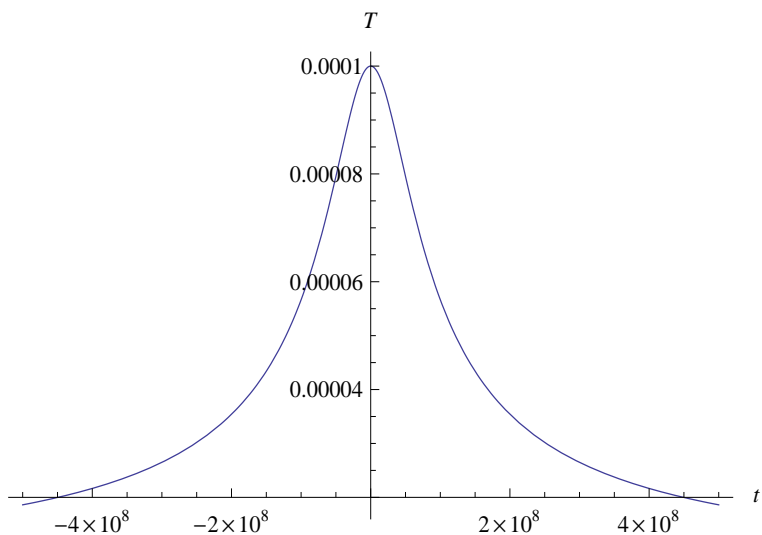}
\caption[]{Figure showing the behavior of the background temperature $T(t)$ in
  the bouncing phase.
}
\label{Ttp}
\end{figure}

With these typical parameter values it is seen that the background
properties of our model encoded in $a(t)$, $H(t)$, $T(t)$, $\rho(t)$
and $p(t)$ show a continuous non-singular bounce. From Fig.~\ref{atp}
it is clear that the scale factor $a(t)$ has a minimum at
$t=0$. Before $t=0$ it is seen that $a(t)$ is decreasing specifying a
contracting universe which smoothly transforms into an expanding
universe after the bounce.  In Fig.~\ref{Htp} the variation of the
Hubble parameter $H(t)$ across the bounce is depicted. The variation
of temperature of the background near the bouncing point is shown in
Fig.~\ref{Ttp}.
\begin{figure}[ht!]
\centering
\includegraphics[width=10cm, height=7cm]{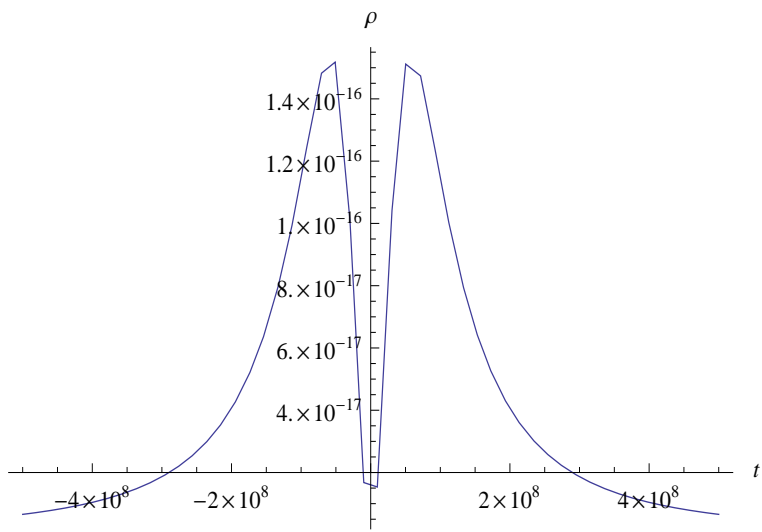}
\caption[]{Figure showing the behavior of the total energy density $\rho(t)$ in
  the bouncing phase.
}
\label{rtp}
\end{figure}
If one compares the variation of energy density with time, as shown in 
Fig.~\ref{rtp}, with the variation in temperature near the bouncing
point one notices that the temperature of the system is the maximum at
the bounce point when the total energy density is the minimum. 
\begin{figure}[ht!]
\centering
\includegraphics[width=10cm, height=7cm]{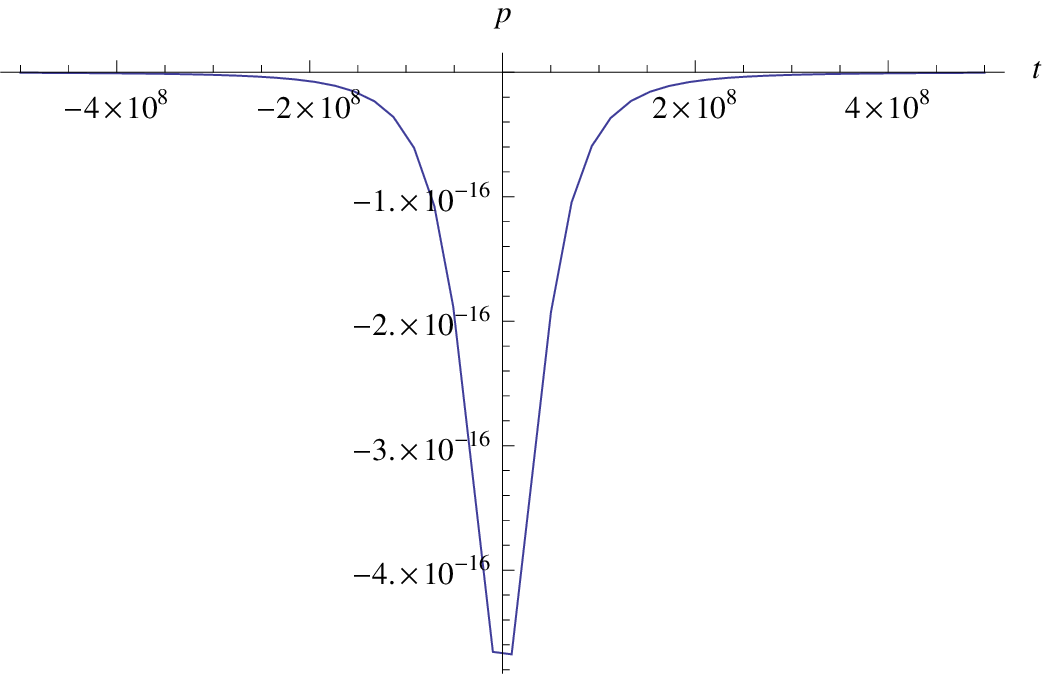}
\caption[]{Figure showing the behavior of pressure $p(t)$ in
  the bouncing phase.
}
\label{ptp}
\end{figure}
This point shows the nature of Lee-Wick particle dominated cosmology
where the Lee-Wick partners contribute to the energy density of the
cosmic plasma with a negative contribution. In Fig.~\ref{ptp} the
pressure of the background plasma, $p(t)$, is plotted against
time. The figure clearly shows that pressure is negative in the
bouncing region, which is a prerequisite of the bouncing model.
\begin{figure}[ht!]
\centering
\includegraphics[width=10cm, height=7cm]{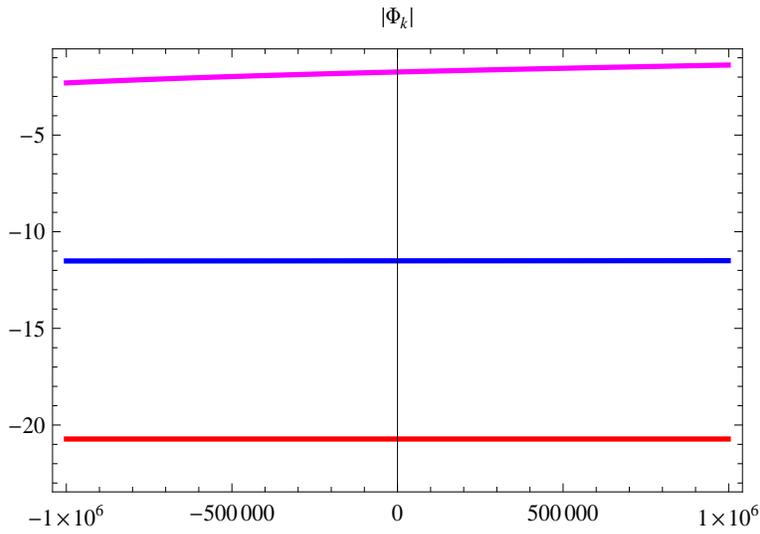}
\caption[]{Figure showing the behavior of the perturbations across the
  bouncing point. In the above figure we have numerically evolved the
  perturbations along the bounce point for the following values of the
  wave number: from top the wave numbers are $k=8\times10^{-7}$,
  $8\times10^{-9}$, $8\times10^{-11}$ respectively. The $y-$axis is
  plotted in logarithmic scale.}
\label{pertp}
\end{figure}

The metric perturbation amplitude $|\Phi_k|$ for three values of the
wave number are plotted in Fig.~\ref{pertp}. Numerical analysis shows
that the perturbations remain almost constant across the bouncing
point. The perturbations smoothly evolves through the bouncing point
and that allows one to match the mode functions in the expanding phase
after the bounce. While comparing these different Fourier modes, one
finds that the perturbation mode with the highest value of $k$ has the
largest amplitude. Thus, it shows explicitly a blue spectrum is
achieved in our model based on thermal initial condition. Therefore,
the numerical result is in agreement with the analytic analysis
performed previously.

\subsection{Thermal fluctuations in the case which does not include CDM}

Here we will discuss the scenario where a bounce can be achieved in
the contracting phase without invoking a CDM component in the cosmic
plasma. The main feature of this model is that the bounce would occur
before all the heaviest Lee-Wick resonances could thermalize. So, some
of the standard model particles will contribute to the radiation
plasma without their Lee-Wick resonances (like the standard radiation
phase) near the bounce. In this case the dominant part of energy
density before $\tau_B^-$ or after $\tau_B^+$ is assumed to be coming
from the standard radiation part, namely from $\tilde c T^4/4$, in
Eq. (\ref{energyd4}). The energy density starts to decrease (or
increase) rapidly in the bouncing phase. Consequently in this scenario
the dominant nature of the radiation plasma would be more like a
conventional radiation fluid with $w_r\approx1/3$. Hence, for the
modes which would leave the Hubble radius in the contracting phase,
much before the bounce takes place, one can consider $w=w_r\approx
1/3$.

The analysis of the IR regime in the other case holds true here as
well as the scenario of onset of radiation era are same in both the
cases. We need to analyze the UV regime separately as now the
background and radiation fluid evolve as
$w\sim w_r\sim1/3$. Following the arguments given in the
previous sub-sections, one can see that the heat capacity of the
plasma during the Hubble-exit of mode $k$ in this case would be
\begin{eqnarray}
C_V(\tau_k)\approx \frac{4\pi}{3}\tilde c(\tau_k)\frac{T_*^3}{H_*^3}
\left(\frac{k}{k_*}\right)^{-3}~,
\end{eqnarray}
and the amplitude of the metric fluctuation with wavenumber $k$ would become
\begin{eqnarray}\label{ic_model2}
 \Phi_H \equiv \Phi_k^-(\tau_H) \simeq
 \frac{(3\pi)^{1/2}}{2^{3/2}M_p^2}
\frac{\sqrt{\tilde c}~T_*^{5/2}}{a_*^{3/2}H_*^{2}}~.
\end{eqnarray}
where we have applied $w=w_r \simeq 1/3$ and consequently $\nu=3$. In
this case, by matching the formula \eqref{ic_model2} and the general
solution \eqref{Phi-pm} in contracting phase, the $D_-$ and $S_-$
modes can be expressed as
\begin{eqnarray}\label{DS-1}
 D_-(k) \simeq \Phi_H\propto k^0~,~~S_-(k) \simeq \left(\frac{{\cal
     H}_{\rm UV}}{k}\right)^3 \Phi_H\propto \frac{1}{k^3}~,
\end{eqnarray}
which show that the contribution of the $D_-$ mode to the $D_+$ mode
in the expanding phase is scale-free whereas the contribution of the
$S_-$ mode for the same is proportional to $1/k$ (following
Eq. (\ref{matching})). Thus, if the $S_-$ mode (being the growing mode
in the contracting phase) contributes dominantly to the $D_+$ mode in
the expanding phase, then we get the power spectrum as follows,
\begin{eqnarray}
 P_\Phi = \frac{k^3}{2\pi^2} |D_+|^2
 = \frac{3 \tilde{c} a_B^2H_{\rm UV}^2T_* }{4\pi M_p^2 a_*^2H_*}\left(\frac{k}{k_*}\right)~,
\end{eqnarray}
which is also a blue-tilted spectrum as before with a spectral index
\begin{eqnarray}\label{n_Phi_2}
 n_\Phi \equiv 1+\frac{d\ln P_\Phi}{d\ln k} = 2~.
\end{eqnarray}
Note that, if the $D_+$ mode in the expanding phase is mainly
inherited from the $D_-$ mode in contracting phase for some reason,
the corresponding power spectrum is proportional to $k^3$. 

Thus from the above analysis we see that here the spectrum remains
equally blue tilted in both IR and UV regime. In the other case, where
a CDM component is added, the spectrum is more blue titled in the UV
regime than that of IR. Therefore, it is not difficult to distinguish
these two models observationally.

\section{Conclusion}

In this article an attempt has been made to apply the esoteric topic
of Lee-Wick thermodynamics in the physics of the early universe
undergoing a non-singular bounce. Bouncing cosmologies try to evade
the Big Bang singularity which plagues most of the recent cosmological
models. Moreover the bouncing universe model presented in the article
do not require inflation to solve the standard cosmological problems
as an inflation-like scenario is built into the bouncing models.

If one includes Lee-Wick partners of standard fields in the theory
then it naturally leads to a radiation dominated phase where the
partners of some of the standard fields contribute with a negative
sign in the effective energy density of the universe. This negative
radiation can lead to a bouncing phase of the universe when the total
energy density reaches zero. There can be two options to produce the
bounce. The first option includes a dark matter component which helps
to produce a negative pressure near the bounce point\footnote{It helps
  to evade the null energy condition near the bounce point.}. A brief
numerical analysis of the background evolution and the evolution of
the perturbations across the bouncing point for this option is
presented in the article.The other option does not require any dark
matter component but it requires that the temperature of the universe
near the bounce point to be less than the masses of some of the
Lee-Wick partners of the standard fields.

It has been shown that for the bounce mechanism to succeed one has to
have the Lee-Wick partners of most of the chiral fermions and massless
gauge bosons to be heavier than a mass scale. This scale depends on
the bounce energy scale. This fact gives us a hint about the mass of
the illusive Lee-Wick partners.

In the bouncing universe scenarios presented in this article, the
relevant perturbation modes leaves the Hubble radius while the
universe is contracting as the Hubble radius starts to shrink. These
modes lie outside the Hubble radius for a brief time. After the bounce
is over and the Hubble radius starts to increase these modes starts to
re-enter. In this way the bouncing universe scenarios can mimic the
effects of inflationary universe models. The bouncing universe models
naturally sets a wave number cut-off, in our case it is called $k_{\rm
  UV}$. In the Lee-Wick thermal radiation induced bouncing models
discussed here we have calculated the power spectrum of the metric
perturbation $\Phi$. In both the cases one gets a blue-tilted power
spectrum. Therefore, the primordial power spectrum induced by thermal
fluctuations in contracting phase is hardly to explain the CMB data
since cosmological observations have proven a nearly scale-invariant
power spectrum. However, this problem can be generally circumvented by
introducing another light scalar field during matter contracting
phase, which is known as the bounce curvaton
mechanism \cite{Cai:2011zx, Cai:2011ci}. 

The theory with Lee-Wick partners are still in their infancy as they
were neglected for a long time because there were other interesting
scenarios in particle physics which could address the most pressing
problems, as the hierarchy problem and other related problems,
there. At the present moment Lee-Wick standard model and Lee-Wick
thermodynamics related works are also trying to address the problems
plaguing particle physics and cosmology. In this regard the present
article discusses about an interesting property of the probable early
universe utilizing the properties of the Lee-Wick theories.

\section*{Acknowledgments}
We thank Robert Brandenberger and Guy Moore for helpful comments on
the manuscript. SD would like to thank Archisman Ghosh for useful
discussions.  The work of YFC is supported in part by an NSERC
Discovery Grant.

\appendix

\section{Transfer relation between the modes solutions in the contracting 
and expanding phases}
\label{transfer}

Here we will give a brief description of how to match the mode
functions of the two phases, contracting and expanding, at the
non-singular bouncing point by making use of the analysis of
\cite{Deruelle:1995kd}. In \cite{Deruelle:1995kd}, the matching
conditions of modes have been extensively derived in a situation where
the stress-energy tensor undergoes a finite discontinuity at a phase
transition. If this sudden change in the stress-energy tensor is due
to a sudden change in the equation of state $w$, then the energy
density remains constant on the hypersurface $\Sigma$ of phase
transition. In such a situation the two matching conditions in
conformal Newtonian gauge (longitudinal gauge) read
\begin{eqnarray}
\Phi^{\pm}_k&=&0,\nonumber\\
\hat{v}^{\pm}_k&=&\left[v_k-\frac{k^2}{3}(\mathcal{H}^\prime-\mathcal{H}^2)^{-1}\Phi_k\right]_{\pm}=0~,
\end{eqnarray}
where $v$ is the Mukhanov-Sasaki variable which can be written in
terms of $\Phi$ as
\begin{eqnarray}
v=\Phi+\frac23\frac{H^{-1}\dot\Phi+\Phi}{(1+w)}=\Phi+\frac{\mathcal{H}}{\mathcal{H}^2-\mathcal{H}^\prime}(\Phi^\prime+\mathcal{H}\Phi),
\label{ms-variable}
\end{eqnarray}
Using $\tau^{-\nu}=\frac{\mathcal{H}}{a^2}$ the first matching
condition yields
\begin{eqnarray}
D_+=D_-+\frac{S_-\mathcal{H}_--S_+\mathcal{H}_+}{a^2},
\label{first-cond}
\end{eqnarray}
where we have used the solutions for $\Phi^\pm_k$ given in
Eq. (\ref{Phi-pm}).  On the other hand the second matching condition
gives
\begin{eqnarray}
&&\frac{1}{\mathcal{H}_-^\prime-\mathcal{H}_-^2}\left[\left(\mathcal{H}_-^2+\frac{k^2}{3}\right)D_-+\frac{S_-\mathcal{H}_-}{a^2}\left(\mathcal{H}_-^\prime-\mathcal{H}_-^2+\frac{k^2}{3}\right)\right]=\nonumber\\
&&\frac{1}{\mathcal{H}_+^\prime-\mathcal{H}_+^2}\left[\left(\mathcal{H}_+^2+\frac{k^2}{3}\right)D_++\frac{S_+\mathcal{H}_+}{a^2}\left(\mathcal{H}_+^\prime-\mathcal{H}_+^2+\frac{k^2}{3}\right)\right].
\label{second-cond}
\end{eqnarray}
Using the first relation given in Eq.~(\ref{first-cond}) in the above
equation one gets
\begin{eqnarray}
\frac{2\mathcal{H}_+^\prime-\mathcal{H}_+^2}{\mathcal{H}_+^\prime-\mathcal{H}_+^2}D_+&=&\left[\frac{2\mathcal{H}_-^\prime-\mathcal{H}_-^2}{\mathcal{H}_-^\prime-\mathcal{H}_-^2}+\frac{k^2}{3}\left(\frac{1}{\mathcal{H}_-^\prime-\mathcal{H}_-^2}-\frac{1}{\mathcal{H}_+^\prime-\mathcal{H}_+^2}\right)\right]D_-\nonumber\\
&&-\frac{S_-\mathcal{H}_-}{a^2}\left(\frac{k^2}{3}\right)\left(\frac{1}{\mathcal{H}_-^\prime-\mathcal{H}_-^2}+\frac{1}{\mathcal{H}_+^\prime-\mathcal{H}_+^2}\right).
\end{eqnarray}
If the universe goes through a symmetric evolution around the bounce
then the above equation leads to
\begin{eqnarray}
D_+=AD_-+Bk^2S_-\,,
\label{d-plus}
\end{eqnarray}
where $A$ and $B$ are constants of $\mathcal{O}(1)$.



\begin{thebibliography}{999}

\bibitem{Lee:1969fy}
  T.~D.~Lee and G.~C.~Wick,
  Nucl.\ Phys.\ B {\bf 9}, 209 (1969).

\bibitem{Lee:1970iw}
  T.~D.~Lee and G.~C.~Wick,
  Phys.\ Rev.\ D {\bf 2}, 1033 (1970).

\bibitem{Grinstein:2007mp}
  B.~Grinstein, D.~O'Connell and M.~B.~Wise,
  Phys.\ Rev.\ D {\bf 77}, 025012 (2008)
  [arXiv:0704.1845 [hep-ph]].

\bibitem{Cline:2003gs} 
  J.~M.~Cline, S.~Jeon and G.~D.~Moore,
  Phys.\ Rev.\ D {\bf 70}, 043543 (2004)
  [hep-ph/0311312].

\bibitem{Carone:2008bs}
  C.~D.~Carone and R.~F.~Lebed,
  Phys.\ Lett.\  B {\bf 668}, 221 (2008)
  [arXiv:0806.4555 [hep-ph]].


\bibitem{Carone1}
 C.~D.~Carone and R.~F.~Lebed,
  JHEP {\bf 0901}, 043 (2009)
  [arXiv:0811.4150 [hep-ph]].


\bibitem{Carone2}
C.~D.~Carone,
  Phys.\ Lett.\  B {\bf 677}, 306 (2009)
  [arXiv:0904.2359 [hep-ph]].


\bibitem{Carone3}
C.~D.~Carone and R.~Primulando,
  Phys.\ Rev.\  D {\bf 80}, 055020 (2009)
  [arXiv:0908.0342 [hep-ph]].

\bibitem{Alvarez:2011ah}
  E.~Alvarez, E.~C.~Leskow and J.~Zurita,
  Phys.\ Rev.\ D {\bf 83}, 115024 (2011)
  [arXiv:1104.3496 [hep-ph]].

\bibitem{Krauss:2007bz}
  F.~Krauss, T.~E.~J.~Underwood and R.~Zwicky,
  Phys.\ Rev.\ D {\bf 77}, 015012 (2008)
  [Erratum-ibid.\ D {\bf 83}, 019902 (2011)]
  [arXiv:0709.4054 [hep-ph]].

\bibitem{Figy:2011yu}
  T.~Figy and R.~Zwicky,
  JHEP {\bf 1110}, 145 (2011)
  [arXiv:1108.3765 [hep-ph]].


\bibitem{Cai:2008qw}
  Y.~-F.~Cai, T.~-t.~Qiu, R.~Brandenberger and X.~-m.~Zhang,
  Phys.\ Rev.\ D {\bf 80}, 023511 (2009)
  [arXiv:0810.4677 [hep-th]].


\bibitem{Wands:1998yp}
  D.~Wands,
  Phys.\ Rev.\ D {\bf 60}, 023507 (1999)
  [gr-qc/9809062].

\bibitem{Finelli:2001sr}
  F.~Finelli and R.~Brandenberger,
  Phys.\ Rev.\ D {\bf 65}, 103522 (2002)
  [hep-th/0112249].


\bibitem{Karouby:2010wt}
  J.~Karouby and R.~Brandenberger,
  Phys.\ Rev.\ D {\bf 82}, 063532 (2010)
  [arXiv:1004.4947 [hep-th]].

\bibitem{Karouby:2011wj}
  J.~Karouby, T.~Qiu and R.~Brandenberger,
  Phys.\ Rev.\ D {\bf 84}, 043505 (2011)
  [arXiv:1104.3193 [hep-th]].


\bibitem{Fornal:2009xc}
  B.~Fornal, B.~Grinstein and M.~B.~Wise,
  Phys.\ Lett.\ B {\bf 674}, 330 (2009)
  [arXiv:0902.1585 [hep-th]].

\bibitem{Bhattacharya:2011bb}
  K.~Bhattacharya and S.~Das,
  Phys.\ Rev.\ D {\bf 84}, 045023 (2011)
  [arXiv:1108.0483 [hep-ph]].

\bibitem{Wise:2009mi}
  M.~B.~Wise,
  Int.\ J.\ Mod.\ Phys.\ A {\bf 25}, 587 (2010)
  [arXiv:0908.3872 [hep-ph]].

\bibitem{Bhattacharya:2012te}
  K.~Bhattacharya and S.~Das,
  Phys.\ Rev.\ D {\bf 86}, 025009 (2012)
  [arXiv:1203.1109 [hep-ph]].


\bibitem{Mukhanov:1990me}
  V.~F.~Mukhanov, H.~A.~Feldman and R.~H.~Brandenberger,
  Phys.\ Rept.\  {\bf 215}, 203 (1992).  


\bibitem{Cai:2012va}
  Y.~-F.~Cai, D.~A.~Easson and R.~Brandenberger,
  JCAP {\bf 1208}, 020 (2012)  [arXiv:1206.2382 [hep-th]].

\bibitem{Cai:2009rd}
  Y.~-F.~Cai, W.~Xue, R.~Brandenberger and X.~-M.~Zhang,
  JCAP {\bf 0906}, 037 (2009)  [arXiv:0903.4938 [hep-th]].

\bibitem{Hwang:1991an}
  J.~-c.~Hwang and E.~T.~Vishniac,
  Astrophys.\ J.\  {\bf 382}, 363 (1991).

\bibitem{Deruelle:1995kd}
  N.~Deruelle and V.~F.~Mukhanov,
  Phys.\ Rev.\ D {\bf 52}, 5549 (1995)  [gr-qc/9503050].

\bibitem{Cai:2007zv}
  Y.~-F.~Cai, T.~Qiu, R.~Brandenberger, Y.~-S.~Piao and X.~Zhang,
  JCAP {\bf 0803}, 013 (2008)  [arXiv:0711.2187 [hep-th]].

\bibitem{Cai:2008ed}
  Y.~-F.~Cai and X.~Zhang,
  JCAP {\bf 0906}, 003 (2009)  [arXiv:0808.2551 [astro-ph]].

\bibitem{Martin:2003sf} 
  J.~Martin and P.~Peter,
  Phys.\ Rev.\ D {\bf 68}, 103517 (2003)
  [hep-th/0307077].

\bibitem{Cai:2011tc}
  Y.~-F.~Cai, S.~-H.~Chen, J.~B.~Dent, S.~Dutta and E.~N.~Saridakis,
  Class.\ Quant.\ Grav.\  {\bf 28}, 215011 (2011)  [arXiv:1104.4349
  [astro-ph.CO]].  

\bibitem{Martin:2000xs}
  J.~Martin and R.~H.~Brandenberger,
  Phys.\ Rev.\ D {\bf 63}, 123501 (2001)
  [hep-th/0005209].  

\bibitem{Brandenberger:2000wr}
  R.~H.~Brandenberger and J.~Martin,
  Mod.\ Phys.\ Lett.\ A {\bf 16}, 999 (2001)
  [astro-ph/0005432].  

\bibitem{Brandenberger:1999sw}
  R.~H.~Brandenberger,
  hep-ph/9910410.  


\bibitem{Cai:2011zx}
  Y.~-F.~Cai, R.~Brandenberger and X.~Zhang,
  JCAP {\bf 1103}, 003 (2011)
  [arXiv:1101.0822 [hep-th]].  

\bibitem{Cai:2011ci}
  Y.~-F.~Cai, R.~Brandenberger and X.~Zhang,
  Phys.\ Lett.\ B {\bf 703}, 25 (2011)
  [arXiv:1105.4286 [hep-th]].  


\end{thebibliography}
\end{document}